\documentstyle[prl,aps,epsf,floats]{revtex}

\begin{document}
\draft

\twocolumn[\hsize\textwidth\columnwidth\hsize\csname @twocolumnfalse\endcsname

\title{Spin accumulation in degenerate semiconductors near modified Schottky
contact with ferromagnets}
\author{V.V. Osipov$^{1,2}$ and  A.M. Bratkovsky$^{1}$  } 

\address{
${^1}$Hewlett-Packard Laboratories, 1501 Page Mill Road, 1L, Palo Alto, California 94304
}
\address{
${^2}$New Physics Devices, NASA Ames, Moffet Field, California 94035}
\date{April 18, 2005}
\maketitle

\begin{abstract}
We study spin transport in forward and reverse biased junctions
between a ferromagnetic metal and a  degenerate semiconductor with a
$\delta -$doped layer near the 
interface at relatively low temperatures.  
We show that spin polarization of electrons in the semiconductor,
$P_n$, near the interface increases both  
with the forward and reverse current and reaches saturation at certain relatively large
current while the spin injection coefficient, $\Gamma$, increases with reverse current and
decreases with the forward current. We analyze the condition for
efficient spin polarization  
of electrons in degenerate semiconductor near interface with ferromagnet. We compare the
accumulation of spin polarized electrons in degenerate semiconductors
at low temperatures with that in nondegenerate semiconductors at relatively high, room
temperatures. 
\pacs{72.25.Hg, 72.25.Mk}
\end{abstract}

\vskip2pc]

\narrowtext

\section{Introduction}

The idea of novel solid state electronic devices using a electron spin has
given rise to the new field of spintronics \cite{Wolf,zuticRev04}. Among
practically important spintronic effects are the giant magnetoresistance in
magnetic multilayers and the tunnel ferromagnet-insulator-ferromagnet
(FM-I-FM) structures \cite{GMR,Slon,Brat}. Of particular interest is
injection of spin-polarized electrons into semiconductors because of large
spin relaxation time \cite{SpinM} and a prospect of using this phenomena for
the next generation of high-speed low-power electronic devices \cite
{Datta,Hot,BO} and quantum computing \cite{Wolf}. Relatively efficient spin
injection into nonmagnetic semiconductors (S) has been demonstrated at low
temperatures in ferromagnet-semiconductor heterostructures both with
metallic ferromagnets \cite{Ferro,Jonk,Ohno} and magnetic semiconductors 
\cite{MSemi} as the spin sources. Theoretical aspects of the spin injection
have been studied in Refs.~\cite
{Aron,Mark,Son,Sch00,Flat,Alb,Her,Rash,Fert,Hu,Flat1,OB,BOB}.

There are several fundamentally different types of FM-S junctions with the
energy band diagrams shown in Fig.~1. The band diagrams depend on electron
affinity of a semiconductor, $\chi _{S}$, and a work function of a
ferromagnet, $\chi _{F}$, electron density in a semiconductor, $n$, and a
density of surface states at the FM-S interface \cite{sze}. Usually, a
depleted layer and a high Schottky potential barrier form in S near
metal-semiconductor junction, Fig.~1a,b, at $\chi _{S}>\chi _{F}$ and even
when $\chi _{S}>\chi _{F},$ due to the presence of surface states on the
FM-S interface \cite{sze}. In some systems with $\chi _{S}>\chi _{F},$ a
layer with accumulated electrons can form in S near the FM-S interface,
Figs.~1c,d. Such a rare situation is probably realized in Fe-InAs junctions
studied in Ref. \cite{Ohno}. The barrier height in the usual situation
(Figs.~1a,b) is equal to $\Delta \simeq 0.5-0.8$ eV for GaAs and Si in
contacts with practically all metals, including Fe, Ni, and Co \cite
{sze,Jonk}. The barrier width, i.e. the Schottky depleted layer width, is
large, $l\gtrsim 30$ nm, for doping donor concentration $N_{d}\lesssim
10^{17}$cm$^{-3}$. The injection of spin polarized electrons from FM into S
corresponds to a reverse current in the Schottky contact, when positive
voltage is applied to $n-$S region. The current in reverse-biased FM-S
Schottky contacts is saturated and usually negligible due to such large
barrier thickness and height, $l$\ and $\Delta $\ \cite{sze}. Therefore, a
thin heavily doped $n^{+}-$S layer between FM metal and S should used to
increase the reverse current determining the spin-injection \cite{Alb,OB}.
This layer drastically reduces the thickness of the barrier, and increases
its tunneling transparency \cite{sze,OB}. Thus, an efficient spin injection
has been observed in FM-S junctions with a thin $n^{+}-$layer \cite{Jonk}.
\begin{figure}[t]
\epsfxsize=3.6in 
\epsffile{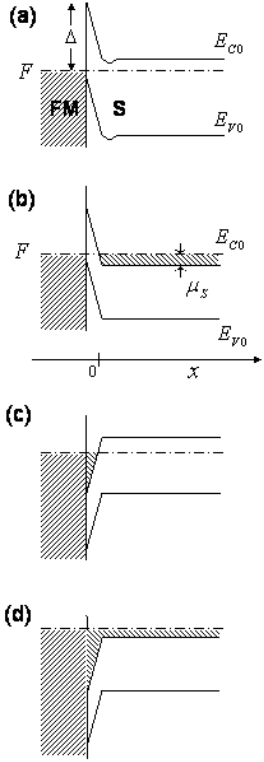}

\caption{Energy diagrams of modified ferromagnet-semiconductor (FM-S)
junctions at equilibrium: modified Schottky
contact of a FM metal with nondegenerate (a) and degenerate (b)
semiconductors with a depletion layer in S near the interface. 
The Schottky contacts are modified by  highly doped very thin
semiconductor $\delta -$doped layer between FM and S, $\Delta$
is the Schottky barrier height, $\mu_S$ the Fermi energy in the
degenerate S. 
Band diagrams (c) and (d) are for modified ohmic contacts of a FM
metal with nondegenerate (c) and
degenerate (d) semiconductors with an accumulation layer in S.
}
\label{fig:fig1}
\end{figure}

In forward-biased FM-S Schottky contacts without the thin $n^{+}-$layer,
current can reach a large value only at a bias voltage $V$ close to $\Delta
/q$, where $q$ is the elementary charge \cite{sze}. Realization of the spin
accumulation in S due to such thermionic emission currents is problematic.
Indeed, electrons in FM with energy $F+\Delta $ well above the Fermi level $%
F $ are weakly spin polarized.

The energy band structure of FM-S junctions, their spin-selective and
nonlinear properties have not been actually considered in majority of
theoretical works on spin injection \cite
{Aron,Mark,Son,Sch00,Flat,Alb,Her,Rash,Fert,Hu,Flat1}. Authors of these
prior works have developed a {\em linear }theory of spin injection
describing the spin-selective properties of FM-S junctions by various, often
contradictory, boundary conditions at the FM-S interface. For example,
Aronov and Pikus assumed that a {\em spin injection coefficient} (spin
polarization of current in FM-S junctions) $\Gamma =\left( J_{\uparrow
}-J_{\downarrow }\right) /J$ at the FM-S interface is a constant, equal to
that in the FM metal, and studied spin accumulation in semiconductors
considering spin diffusion and drift in applied electric field \cite{Aron}.
The authors of Refs. \cite{Mark,Son,Sch00,Flat,Alb} assumed a continuity of
both the currents and the electrochemical potentials for both spins\ and
found that a spin polarization of injected electrons depends on a ratio of
conductivities of a FM and S (the so-called \textquotedblleft conductivity
mismatch\textquotedblright\ problem). At the same time, the authors of
Refs.~ \cite{Her,Rash,Fert,Hu,Flat1} have asserted that the spin injection
becomes appreciable when the electrochemical potentials have a substantial
discontinuity at the interface (produced by e.g. a tunnel barrier \cite{Rash}%
). However, they described this effect by the unknown constants,
spin-selective interface conductances $G_{\sigma }$, which cannot be found
within those theories. In fact, we have shown before that\ the parameters $%
G_{\sigma }$ are not constant and can strongly depend on the applied bias
voltage \cite{BO,OB}.

In our earlier works \cite{BO,OB,BOB} we have studied the nonlinear spin
injection in nondegenerate semiconductors near modified FM-S Schottky
contacts with $\delta -$doped layer, Fig.1a, at room temperature and showed
that the assumptions made in Refs.~\cite{Mark,Son,Sch00,Flat,Alb} are not
valid at least in that case. Here we derive the boundary conditions and
study nonlinear spin injection in degenerate semiconductors near reverse-
and forward-biased FM-S Schottky contacts with an ultrathin heavily doped
semiconductor layer ($\delta -$doped layer) between FM and S, Fig.~1b. In
degenerate semiconductors, unlike in nondegenerate semiconductors studied in
Refs. \cite{OB,BOB}, the spin injection can occur at any (low) temperatures.
We consider below the case when the temperature $T\ll \mu _{S}$, where $\mu
_{S}=(F-E_{c0})$ is the Fermi energy of equilibrium electrons in S, where $%
E_{c0}$ is the bottom of the conduction band in equilibrium, Fig.~1, and $T$
is the temperature in units of $k_{B}=1$.

\section{Spin tunneling through thin $\delta -$doped barrier at FM-S interface}

We assume that the donor concentration, $N_{d}^{+}$, and thicknesses, $l$,
of the $\delta -$doped layer satisfy the conditions $N_{d}^{+}l^{2}q^{2}%
\simeq 2\varepsilon \varepsilon _{0}\Delta $ and $l\lesssim l_{0},$ where $%
l_{0}=\sqrt{\hbar ^{2}/(2m_{\ast }\Delta )}$ is a typical tunneling length ($%
l_{0}\lesssim 2$ nm for $N_{d}^{+}\sim 10^{20}$cm$^{-3}$). The energy band
diagram of such a FM-S junction includes a potential $\delta -$spike of the
height $\Delta $ and the thickness $l$ shown in Fig.~1b. We assume the
elastic coherent tunneling through this $\delta -$layer, so that the energy $%
E$, spin $\sigma ,$ and the component of the wave vector $\vec{k}$ parallel
to the interface, $\vec{k}_{\parallel }$, are conserved. In this case the
tunneling current density of electrons with spin $\sigma =\uparrow
,\downarrow $ near the FM-S junction containing the $\delta -$doped layer
(Fig.~1) can be written as \cite{Duke,Brat,OB,BOB}

\begin{eqnarray}
J_{\sigma 0} &=&\frac{-q}{(2\pi )^{3}}\int d^{3}k[f(E_{k\sigma }-F_{\sigma
0}^{f})-f(E_{k\sigma }-F_{\sigma 0}^{S})]v_{\sigma x}T_{k\sigma }  \nonumber
\\
&=&\frac{q}{h}\int dE[f(E-F_{\sigma 0}^{S})-f(E-F_{\sigma 0}^{f})]\int \frac{%
d^{2}k_{\parallel }}{(2\pi )^{2}}T_{\sigma },  \label{GEq}
\end{eqnarray}
where $T_{k\sigma }$ is the transmission probability, $f(E)$ the Fermi
function, $v_{\sigma x}$ the $x-$component of velocity\ $v_{\sigma }=\hbar
^{-1}|\nabla _{k}E_{k\sigma }|$ of electrons with the wave vector $\vec{k}$
and spin $\sigma $\ in the ferromagnet, the integration includes a summation
with respect to a band index. Importantly, one needs to account for a strong 
{\em spin accumulation} in the semiconductor. Therefore, we use the {\em %
nonequilibrium} Fermi levels, $F_{\sigma 0}^{f}$ and $F_{\sigma 0}^{S}$ for
electrons with spin $\sigma =\uparrow (\downarrow )$ in the FM\ metal and
the semiconductor, respectively, near the interface, $x=0$. In particular,
the local electron density with spin $\sigma $\ in the degenerate
semiconductor at the FM-S junction\ at low temperatures is given by 
\begin{eqnarray}
n_{\sigma 0} &=&\frac{2^{1/2}m_{\ast }^{3/2}M_{c}}{3\pi ^{2}\hbar ^{3}}%
(F_{\sigma 0}^{S}-E_{c})^{3/2}  \nonumber \\
&=&\frac{n}{2\mu _{S}^{3/2}}(F_{\sigma 0}^{S}-E_{c})^{3/2}=\frac{n}{2}\left(
1+\frac{\Delta F_{\sigma 0}^{S}}{\mu _{S}}\right) ^{3/2},  \label{nc}
\end{eqnarray}
where $M_{c}$ the number of effective minima of the semiconductor conduction
band; $E_{c0}$ and $E_{c}=E_{c0}+qV$\ are the bottom of conduction band in S
at equilibrium and at the bias voltage $V$, $\ \mu _{S}=F-E_{c0}$ is the
equilibrium Fermi {\em energy} of the electrons in the semiconductor bulk,
with $F$ the Fermi level in FM metal bulk, $F_{\sigma 0}^{S}$ is the quasi
Fermi level in S near the interface (point $x=0,$ Fig.~1), $\Delta F_{\sigma
0}^{S}=F_{\sigma 0}^{S}-E_{c}=F_{\sigma 0}^{S}-E_{c0}-qV,$ 
\mbox{$\vert$}%
$\Delta F_{\sigma 0}^{S}|<\mu _{S},$ $n$ and $m_{\ast }$ are the
concentration and effective mass of electrons in S. We note that $V>0$ and
current $J>0$ in forward-biased FM-S junctions , i.e. $J$ flows in $x-$%
direction from FM to S when $V>0$ (usual convention \cite{sze}), and $V<0$
and $J<0$ in reverse biased junctions. The current (\ref{GEq}) should
generally be evaluated numerically for a complex band structure $E_{k\sigma
} $\cite{stefano99}. The analytical expressions for $T_{\sigma
}(E,k_{\parallel })$\ can be obtained in an effective mass approximation, $%
\hbar k_{\sigma }=m_{\sigma }v_{\sigma }$ where $v_{\sigma }$ is the
velocity of electrons in the FM with spin $\sigma $. This applies to
\textquotedblleft fast\textquotedblright\ free-like d-electrons in elemental
ferromagnets \cite{Stearns,Brat}. Approximating the $\delta -$barrier by a
triangular shape, we find 
\begin{eqnarray}
T_{\sigma } &=&{\frac{16\alpha m_{\sigma }m_{\ast }k_{\sigma x}k_{x}}{{%
m_{\ast }^{2}k_{\sigma x}^{2}+m_{\sigma }^{2}}\kappa ^{2}}}e^{-\eta \kappa l}
\nonumber \\
&=&\frac{16\alpha v_{\sigma x}v_{x}}{v_{\sigma x}^{2}+v_{tx}^{2}}e^{-\eta
\kappa l},  \label{ts}
\end{eqnarray}
where $\kappa =(2m_{\ast }/\hbar ^{2})^{1/2}(\Delta +F-E+E_{\parallel
})^{3/2}/(\Delta -qV)$, $E_{\parallel }=\hbar ^{2}k_{\parallel
}^{2}/2m_{\ast }$, $v_{x}=\sqrt{2(E-E_{c}-E_{\parallel })/m_{\ast }}$ is the 
$x-$component of the velocity of electrons in S, $\hbar k_{x}=v_{x}m_{\ast }$%
, $v_{t}=\hbar \kappa /m_{\ast }$ the \textquotedblleft
tunneling\textquotedblright\ velocity, $\alpha =\pi (\kappa l)^{1/3}\left[
3^{1/3}\Gamma ^{2}\left( \frac{2}{3}\right) \right] ^{-1}\simeq 1.2(\kappa
l)^{1/3},$ $\eta =4/3$ (for comparison, for a rectangular barrier $\alpha =1$
and $\eta =2$), $\theta (x)=1$ for $x>0,$ and zero otherwise. The
preexponential factor in Eq.~(\ref{ts}) takes into account a mismatch
between effective mass, $m_{\sigma }$ and $m_{\ast }$, and velocities, $%
v_{\sigma x}$ and $v_{x}$, of electrons in the FM and the S. Obviously, only
the states with $E_{k\sigma }>E_{c}$ are available for transport.

We obtain the following expression for the current at the temperature $T\ll
\mu _{S}$ with the use of Eqs.~(\ref{GEq})\ and (\ref{ts}), noting that the
electron velocity in the semiconductor is singular near $E=E_{c}$ 
\begin{eqnarray}
J_{\sigma 0} &=&\frac{2\pi qm_{\ast }M_{c}}{h^{3}}\biggl[\int_{E_{c}}^{F_{%
\sigma 0}^{S}}dE\int_{0}^{E-E_{c}}dE_{\parallel }T_{\sigma }-  \nonumber \\
&&\int_{E_{c}}^{F_{\sigma 0}^{f}}dE\int_{0}^{E-E_{c}}dE_{\parallel
}T_{\sigma }\biggr],  \label{Js1}
\end{eqnarray}
where the second integral corresponds to electrons tunneling from the metal
into semiconductor that can only take place when $F_{\sigma 0}^{f}>E_{c}.$
As a rule, $v_{\sigma x}$ and $v_{tx}$ are smooth functions over $E$ in
range $E=F\pm \mu _{S}$ of interest to us in comparison with a singular $%
v_{x}$. At not very large bias voltages of interest, $|V|\lesssim \Delta /q$
all factors but $v_{x}$ can be taken outside of integration. We obtain,
therefore, from (\ref{nc}) and (\ref{Js1}) the expression 
\begin{eqnarray}
J_{\sigma 0} &=&\frac{32\pi \alpha _{0}qm_{\ast }M_{c}v_{\sigma }}{%
h^{3}\left( v_{\sigma }^{2}+v_{t}^{2}\right) }e^{-\eta \kappa _{0}l} 
\nonumber \\
&&\times \biggl[\int_{E_{c}}^{F_{\sigma
0}^{S}}dE\int_{0}^{E-E_{c}}dE_{\parallel }v_{x}-  \nonumber \\
&&\theta \left( F_{\sigma 0}^{f}-E_{c}\right) \int_{E_{c}}^{F_{\sigma
0}^{f}}dE\int_{0}^{E-E_{c}}dE_{\parallel }v_{x}\biggr],  \label{eq:JsINT}
\end{eqnarray}
which with use Eq. (\ref{nc}) can be finally written as 
\begin{eqnarray}
J_{\sigma 0} &=&j_{0}d_{\sigma }\biggl[\left( 1+\frac{\Delta F_{\sigma 0}^{S}%
}{\mu _{S}}\right) ^{5/2}-  \nonumber \\
&&\left( 1+\frac{\Delta F_{\sigma 0}^{f}-qV}{\mu _{S}}\right) ^{5/2}\theta
\left( \mu _{S}+\Delta F_{\sigma 0}^{f}-qV\right) \biggr],  \label{eq:Js0}
\end{eqnarray}
where 
\begin{equation}
j_{0}=\frac{4}{5}qnv_{F}^{S}\alpha _{0}\exp (-\eta \kappa _{0}l),  \label{j0}
\end{equation}
\begin{equation}
d_{\sigma }=\frac{v_{F}v_{\sigma 0}}{v_{t0}^{2}+v_{\sigma 0}^{2}}.
\label{ds}
\end{equation}
Here $\kappa _{0}\equiv 1/l_{0}=(2m_{\ast }/\hbar ^{2})^{1/2}(\Delta
-qV)^{1/2}$; $\alpha _{0}=0.96(\kappa _{0}l)^{1/3}$; $v_{F}^{S}=\sqrt{2\mu
_{S}/m_{\ast }}$ is the velocity of electrons in the degenerate
semiconductor, $v_{\sigma 0}=v_{\sigma }$\ the velocity of electrons in the
FM (taken at $E=F$ and $F+qV$ for reverse and forward bias voltages,
respectively, see below), $v_{t0}=\sqrt{2(\Delta -qV)/m_{\ast }}$, and $%
\Delta F_{\sigma 0}^{f}=F_{\sigma 0}^{f}-F$ is the splitting of the quasi
Fermi level $F_{\sigma 0}^{f}$ for nonequilibrium electrons with spin $%
\sigma =\uparrow (\downarrow )$ in FM\ metal. We notice that only spin
factor $d_{\sigma }$determines the dependence of current on materials
parameters of a ferromagnet. The need for a different choice of $v_{\sigma
0} $ for forward and reverse bias voltage is evident from Fig.~2. At forward
bias the electrons tunnel from the semiconductor into the states in the
ferromagnetic metal at $E=F+qV$, so there $v_{\sigma 0}\approx v_{\sigma
}(F+qV).$ At reverse bias voltage $(V<0)$ the electrons tunnel to the
semiconductor in the interval of energies $E_{c0}+qV<E<F.$ In this case the
effective tunnel barrier height is smallest for electrons with energies $%
E\approx F$, so $v_{\sigma 0}\approx v_{\sigma }(F)$. Moreover, at reverse
bias a spatial charge starts to build in semiconductor and a wide barrier
forms at energies $E<E_{c0}.$ Therefore, only electrons in narrow energy
range $E_{c0}-\mu _{S}<E<F$ can tunnel, and the reverse current practically
saturates at $V<-\mu _{S}/q.$
\begin{figure}[t]
\epsfxsize=3.8in 
\epsffile{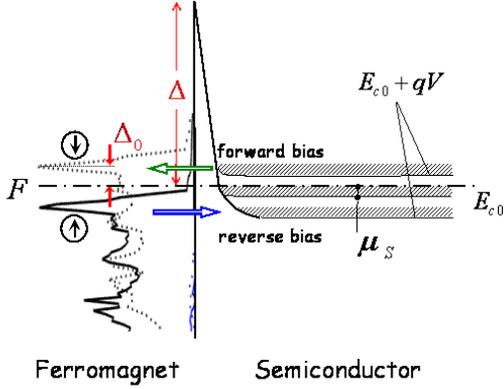}
\caption{Energy diagram of the modified Schottky junction between a FM metal
and an n-type degenerate semiconductor with $\delta -$doped layer at
equilibrium (zero bias, $V=0$)
and at reverse (forward) bias voltage $qV\simeq \Delta _{\downarrow
(\uparrow)}$.  $F$ is the Fermi level in FM, $\Delta $ the barrier height,
$\mu_S$ the Fermi energy in S, $E_{c0}(x)$
the bottom of conduction band of the semiconductor at
equilibrium. Left (right) bold horizontal arrows show a flux of
electrons at forward (reverse) bias voltage. The corresponding states
in FM are dominated by minority (majority) electrons and this may lead
to an accumulation of the spins of the same sign in the semiconductor
at respective forward and reverse bias voltages.
}
\label{fig:fig2}
\end{figure}

Finally, we can present the currents of electrons with spin $\sigma
=\uparrow (\downarrow )$ at the interface in the following useful form: 
\begin{eqnarray}
J_{\uparrow (\downarrow )0} &=&\frac{J_{m0}}{2}(1\pm P_{F})\biggl[\left(
1\pm P_{n}\right) ^{5/3}  \nonumber \\
&&-\left( 1+\frac{\Delta F_{\sigma 0}^{f}-qV}{\mu _{S}}\right) ^{5/2}\theta
(\mu _{S}+\Delta F_{\sigma 0}^{f}-qV)\biggr],  \label{eq:Jsbc}
\end{eqnarray}
where 
\begin{equation}
J_{m0}=(d_{\uparrow }+d_{\downarrow })j_{0}=\frac{4}{5}(d_{\uparrow
}+d_{\downarrow })qnv_{F}^{S}\alpha _{0}\exp (-\eta \kappa _{0}l)  \label{Jm}
\end{equation}
and 
\begin{equation}
P_{F}=\frac{d_{\uparrow }-d_{\downarrow }}{d_{\uparrow }+d_{\downarrow }}=%
\frac{(v_{\uparrow 0}-v_{\downarrow 0})(v_{t0}^{2}-v_{\uparrow
0}v_{\downarrow 0})}{(v_{\uparrow 0}+v_{\downarrow
0})(v_{t0}^{2}+v_{\uparrow 0}v_{\downarrow 0})}.  \label{PF}
\end{equation}
As we see below the value of $P_{F}$ determines maximum spin polarization.

At small bias voltage $qV$ when $\Delta F_{\sigma 0}^{S},$ and $\Delta
F_{\sigma 0}^{f}$ are much smaller than $\mu _{S},$ and by linearizing Eq.~(%
\ref{eq:Js0}) we obtain 
\begin{eqnarray}
J_{\sigma 0} &=&\frac{5}{2}j_{0}d_{\sigma }(\Delta F_{\sigma
0}^{S}+qV-\Delta F_{\sigma 0}^{f})/\mu _{S}  \nonumber \\
&=&G_{\sigma }(\zeta _{\sigma 0}^{S}-\zeta _{\sigma 0}^{f}),  \label{Smal}
\end{eqnarray}
where $\zeta _{\sigma 0}^{S}=F+qV+\Delta F_{\sigma 0}^{S}$ and $\zeta
_{\sigma 0}^{f}=F+\Delta F_{\sigma 0}^{f}$ are the electrochemical
potentials at FM-S interface in the semiconductor and ferromagnet,
respectively, $G_{\sigma }=\frac{5}{2}j_{0}d_{\sigma }/\mu _{S}$ is the
spin-selective interface linear conductance. It is worth noting that if we
were to use the assumption of Refs.~\cite{Mark,Son,Sch00,Flat,Alb} about a
continuity of the electrochemical potentials at FM-S junction, $\zeta
_{\sigma 0}^{S}=\zeta _{\sigma 0}^{f}$, we must have concluded that no
current flows through the junction, $J_{\sigma 0}=0$. We note that the
boundary condition similar to Eq.~(\ref{Smal}) was used in a linear theory
of spin injection in Refs.~ \cite{Her,Rash,Fert,Fert,Flat1}, where $%
G_{\sigma }$ were introduced as some phenomenological constants. Here, we
have found the explicit expressions for the spin conductances $G_{\sigma }$
for the FM-S junction under consideration. Obviously, $G_{\sigma }$, as well
as $\zeta _{\sigma 0}^{S}$ and $\zeta _{\sigma 0}^{f},$ are not universal
and depend on all specific parameters of the junctions, Fig.~1 (cf. Ref.~ 
\cite{OB}). Moreover, the conclusions drawn from the linear approximation
strongly differ from the results of a full nonlinear analysis provided below
(see also Ref.~\cite{OB}).

Importantly, we can neglect the quasi Fermi splitting in FM\ metal compared
to that in the semiconductor because the density of electrons in the FM\
metal is several orders of magnitude larger that in real semiconductors. It
is easy to prove that $\Delta F_{\sigma 0}^{f}\ll qV$ for the currents of
interest to us (see Appendix B), therefore we can simplify the expression (%
\ref{eq:Jsbc}) for tunneling currents of spin-polarized electrons as 
\begin{eqnarray}
J_{\uparrow (\downarrow )0} &=&\frac{J_{m0}}{2}(1\pm P_{F})\biggl[\left(
1\pm P_{n}\right) ^{5/3}  \nonumber \\
&&-\left( 1-\frac{qV}{\mu _{S}}\right) ^{5/2}\theta (\mu _{S}-qV)\biggr].
\label{eq:Jsbc1}
\end{eqnarray}

\section{Injected and extracted spin polarization in degenerate semiconductor%
}

The assumption of elastic coherent tunneling means a continuity of the
currents $J_{\sigma 0}$ of spin-polarized electrons through the FM-S
junction. In this case the FM-S junction can be characterized by the {\em %
spin injection coefficient} $\Gamma $ according to the definition: 
\begin{equation}
\Gamma =(J_{\uparrow 0}-J_{\downarrow 0})/J,  \label{PJ}
\end{equation}
where $J_{\sigma 0}\equiv J_{\sigma }(0)$ are the currents of electrons with 
$\sigma =\uparrow \left( \downarrow \right) $ near the FM-S interface,
Fig.~1. Notice that $\Gamma $ is the spin polarization of a current in the
FM-S junction, therefore we used symbol $P_{J}$ instead of $\Gamma $ in our
earlier papers \cite{BO,OB,BOB}.

The following derivation of bi-spin diffusion applies to both semiconductor
and ferromagnet based on an assumption of {\em quasineutrality} (see
Appendix B). The current $J_{\sigma }$ is given by 
\begin{equation}
J_{\sigma }=\sigma _{\sigma }E+qD_{\sigma }dn_{\sigma }/dx,  \label{JC}
\end{equation}
where\ $\sigma _{\sigma }=q\mu _{\sigma }n_{\sigma }$, $D_{\sigma },$ $\mu
_{\sigma }$ and $n_{\sigma }$\ are the conductivity, the diffusion constant,
the mobility and the density of electrons\ with spin $\sigma =\uparrow
(\downarrow )$, respectively, $E$ the electric field in S or FM. We assume 
{\em quasineutrality}, $n=n_{\uparrow }(x)+n_{\downarrow }(x)={\rm const}$
and later prove that it holds very well indeed (see Appendix A) and a
continuity of the total current, $J=J_{\uparrow }(x)+J_{\downarrow }(x)={\rm %
const}$, so that one has 
\begin{equation}
\delta n_{\uparrow }=n_{\uparrow }-n_{\uparrow }^{0}=-\delta n_{\downarrow },
\label{n-n}
\end{equation}
and for the electric field 
\begin{equation}
E=\frac{J}{\sigma }-\frac{q(D_{\uparrow }-D_{\downarrow })}{\sigma }\frac{%
d\delta n_{\uparrow }}{dx},  \label{Ex}
\end{equation}
where $\sigma =\sigma _{\uparrow }+\sigma _{\downarrow }=q(\mu _{\uparrow
}n_{\uparrow }+\mu _{\downarrow }n_{\downarrow })$ is the total conductivity
of S or FM. Substituting (\ref{Ex}) into (\ref{JC}), we find 
\begin{equation}
J_{\uparrow (\downarrow )}=\frac{\sigma _{\uparrow (\downarrow )}}{\sigma }%
J+q\bar{D}\frac{d\delta n_{\uparrow (\downarrow )}}{dx},  \label{J1}
\end{equation}
where $\bar{D}=\left( \sigma _{\uparrow }D_{\downarrow }+\sigma _{\downarrow
}D_{\uparrow }\right) /\sigma $ is the {\em bi-spin diffusion }constant for
the semiconductor or ferromagnet.

The bi-spin diffusion that appears in the case of degenerate semiconductors
is different in comparison with nondegenerate semiconductor where $D_{\sigma
}$ and $\mu _{\sigma }$ do not depend on spin orientation (see Refs. \cite
{Aron,Flat,OB}). In degenerate semiconductors we need to account for the
density dependence of the diffusion constant. We will assume that the
relaxation time of electron momentum $\tau $\ weakly depends on a quasi-
Fermi level (i.e. on electron density). Therefore, the mobility of electrons 
$\mu _{\sigma }$ in nonmagnetic semiconductors in question weakly depends on
the electron density. In this case we can put $\mu _{\sigma }=\mu $, $\sigma
_{\sigma }=\sigma n_{\sigma }/n,$ and $D_{\sigma }=(1/3)v_{\sigma }^{2}\tau
=D_{0}(2n_{\sigma }/n)^{2/3}$\ at low temperature, where $D_{0}$ and $\sigma 
$\ are the diffusion coefficient in a non-polarized semiconductor and the
total conductivity of the semiconductor, respectively. The account for a
density-dependent diffusion coefficient gives the following expression for
the bi-spin diffusion coefficient: 
\begin{eqnarray}
\bar{D}[n] &=&D_{0}\biggl[\frac{n_{\uparrow }}{n}\left( \frac{2n_{\downarrow
}}{n}\right) ^{2/3}+\frac{n_{\downarrow }}{n}\left( \frac{2n_{\uparrow }}{n}%
\right) ^{2/3}\biggr]  \nonumber \\
&\equiv &D_{0}u(P_{n}).  \label{eq:Dn}
\end{eqnarray}
where 
\begin{equation}
u=\frac{1}{2}\left[ (1+P_{n})(1-P_{n})^{2/3}+(1-P_{n})(1+P_{n})^{2/3}\right]
,  \label{eq:uP}
\end{equation}
and we have introduced the spin polarization of electrons 
\begin{equation}
P_{n}=\frac{n_{\uparrow }-n_{\downarrow }}{n_{\uparrow }+n_{\downarrow }}=%
\frac{2\delta n_{\uparrow }}{n}.  \label{eq:Pn}
\end{equation}
When the polarization is small, $P_{n}\ll 1$ (as is always the case at
distances $x\gtrsim L_{s}$ from the interface),\ the bi-spin diffusion
coefficient has only quadratic corrections to the usual diffusion
coefficient, $\bar{D}\approx D_{0}\left[ 1-(7/9)P_{n}^{2}\right] ,$ and $%
\bar{D}$ is quite close to the diffusion coefficient in a nondegenerate
semiconductor $D_{0}.$

In nonmagnetic semiconductors the electron density $n_{\sigma }$ is
determined by the continuity equation \cite{Aron,Flat} 
\begin{equation}
dJ_{\sigma }/dx=q\delta n_{\sigma }/\tau _{s},  \label{CC}
\end{equation}
where $\delta n_{\sigma }=n_{\sigma }-n/2$, $n$ is the total density of
equilibrium electrons , $\tau _{s}$ is spin-coherence lifetime of electrons
in S. The expression for current (\ref{J1}) now gives 
\begin{equation}
\frac{J}{n}\frac{dn_{\sigma }}{dx}+q\frac{d}{dx}\left( \bar{D}[n]\frac{%
dn_{\sigma }}{dx}\right) =\frac{q\delta n_{\sigma }}{\tau _{s}}.
\label{eq:kinnl}
\end{equation}
We can rewrite this as an equation for the polarization distribution $%
P_{n}(x),$ using\ $\delta n_{\uparrow }=-\delta n_{\downarrow }$ and $%
n_{\uparrow }/n=(1+P_{n})/2,$ as 
\begin{equation}
\frac{J}{J_{s}}\frac{dP_{n}}{d\tilde{x}}+\frac{d}{d\tilde{x}}\left( u\frac{%
dP_{n}}{d\tilde{x}}\right) =P_{n},  \label{eq:kinNLx}
\end{equation}
where $\tilde{x}=x/L_{s}$ is the dimensionless coordinate and 
\begin{equation}
L_{s}=\sqrt{D_{0}\tau _{s}},\qquad J_{S}=\frac{qnD_{0}}{\tau _{s}}=\frac{%
qnL_{s}}{\tau _{s}},  \label{eq:LsJs}
\end{equation}
are the typical spin-diffusion length and the characteristic current
density. It is very convenient to rewrite the spin currents (\ref{J1})\
through $P_{n}$ as 
\begin{equation}
J_{\uparrow (\downarrow )}=\frac{J}{2}(1\pm P_{n})\pm \frac{J_{s}}{2}u\frac{%
dP_{n}}{d\tilde{x}}.  \label{eq:Jsx}
\end{equation}
The spin currents at the interface $x=0$ should be equal to the tunneling
spin currents through FM-S junction given by Eq.~(\ref{eq:Jsbc1}). With the
use $J=J_{\uparrow }(x)+J_{\downarrow }(x)={\rm const}$, this gives the main
boundary condition at the interface 
\begin{eqnarray}
&&\frac{J}{J_{s}}(P_{n0}-P_{F})+u\left( \frac{dP_{n}}{d\tilde{x}}\right)
_{x=0}  \nonumber \\
&=&\frac{J_{m}}{2J_{s}}[(1+P_{n0})^{5/3}-(1-P_{n0})^{5/3}],
\label{eq:bcmain}
\end{eqnarray}
where 
\begin{equation}
J_{m}=J_{m0}(1-P_{F}^{2}),  \label{eq:Im}
\end{equation}
$P_{n0}=P_{n}(x=0)$ the spin polarization next to the interface. It is easy
to see that the equation (\ref{eq:kinNLx}) becomes linear $(u=1)$ away from
the interface where $P_{n}(x)\rightarrow 0$. Therefore, it has an asymptotic
behavior \cite{Aron,Flat,OB} 
\begin{eqnarray}
P_{n}(x) &=&A\exp (-x/L),\qquad \text{when }x\gg L,  \label{eq:nss} \\
L/L_{s} &=&\sqrt{1+(J/2J_{S})^{2}}-J/2J_{S},  \label{eq:LoLs}
\end{eqnarray}
where the coefficient $A$ would have been equal $A=P_{n0}$ for $u=1$ ($%
D_{\sigma }=D_{0}$ case) like in nondegenerate semiconductor \cite{OB}. The
stationary polarization distribution $P_{n}(x)$ is found from equation (\ref
{eq:kinNLx})\ solved with the\ boundary conditions (\ref{eq:bcmain}) and (%
\ref{eq:nss}).

Interestingly, the effect of {\em nonlinearity of the diffusion coefficient}
in degenerate semiconductors, given by the function $u(P_{n})$ in Eq.(\ref
{eq:uP}),\ appears {\em \ to be very small}. This is confirmed by comparing
the solution of (\ref{eq:kinNLx})\ with the case of constant diffusion
coefficient, $u=1$, but we first give simple arguments why this is so.
Nonlinearity could have only been important in the equation (\ref{eq:kinNLx}%
) when the polarization is close to unity, $P_{n0}\approx 1$, so that $u\ll
1.$ At the same, relatively large $P_{n0}$ can only be achieved at a large
current $J\gg J_{S}$ and a large polarization in ferromagnet, $P_{F}\approx
1 $ (i.e. for a half-metallic FM \cite{Brat}), but even in this case $P_{n}$
remains considerably smaller than $P_{F},$ see Fig.~3 and discussion below.
As a result, the polarization dependence of the diffusion changes the
polarization profile $P_{n}(x)$ very little, see Fig.~3 where we compare the
exact polarization profile with that for $\bar{D}=D_{0}.$
\begin{figure}[t]
\epsfxsize=3.4in 
\epsffile{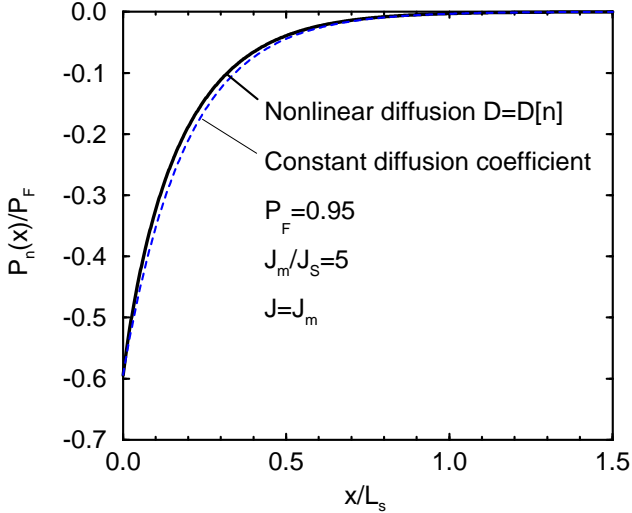}
\caption{Spatial profiles of accumulated polarization with a full
account for non-linear density dependence of diffusion coefficient
($\bar{D}=\bar{D}[n]$, numerical solution) and for constant diffusion
coefficient ($\bar{D}=\bar{D}_0$, analytical solution) for $P_F=0.95$,
$J_m/J_S=5$ and $J=J_m$ (forward bias). Note that in this case the
spin penetration length is $L\approx L_sJ_S/J_m=\frac{1}{5}L_s$ so
that the spin accumulation layer is squeezed towards the interface.
}
\label{fig:fig3}
\end{figure}

We study the current dependence of the polarization in Fig.~4. It
illustrates two important points: (i) the effect of $P_{F},$ the
polarization of injected carriers in a semiconductor and (ii)\ the effect of
having different maximal currents through the structure $J_{m}$ in
comparison with the characteristic current density $J_{S}$. We see that the
difference between the polarization-dependent and independent diffusion
coefficients is minute at all parameters. A small difference is only present
for the spin extraction near maximal current $J\approx J_{m}$ for $%
J_{m}/J_{S}=5$ where the non-linearity in the diffusion coefficient slightly
reduces the extracted polarization. In the opposite case of relatively small
maximal current, $J_{m}/J_{S}=0.2,$ the difference in polarizations is not
discernible at all. The case of $J_{m}/J_{S}\gg 1$ is of most interest to
us, since there the absolute value of the accumulated polarization is
maximal. The overall behavior of the injected/extracted polarization with
the current is similar to the one we found for non-degenerate semiconductors 
\cite{BO,BOB}.

Since we have determined that the density dependence of the diffusion
coefficient in a semiconductor has little effect, the solution of the
kinetic equation (\ref{eq:kinNLx}) reduces to (\ref{eq:nss}), (\ref{eq:LoLs}%
), where the prefactor $A=P_{n0}$. The boundary condition (\ref{eq:bcmain})
then simplifies to 
\begin{equation}
P_{n0}\frac{L}{L_{s}}+\frac{J_{m}}{2J_{S}}\left[
(1+P_{n0})^{5/3}-(1-P_{n0})^{5/3}\right] =-P_{F}\frac{J}{J_{S}}.
\label{eq:bclin}
\end{equation}
This is similar to the case of nondegenerate semiconductor, the difference
being the term in square brackets, which is specific for the degenerate
semiconductor.
\begin{figure}[t]
\epsfxsize=3.4in 
\epsffile{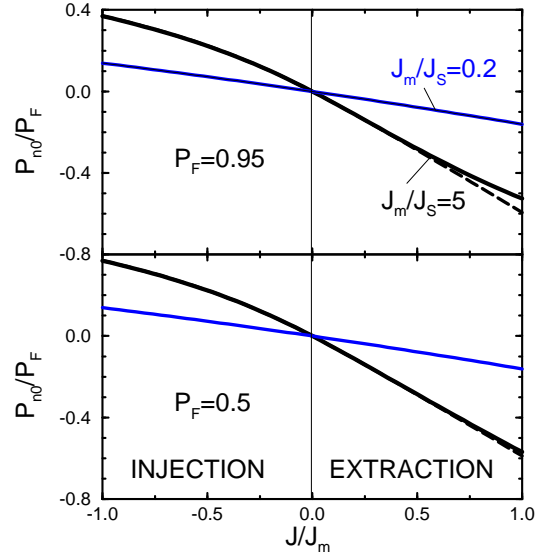}
\caption{Current dependence of spin polarization near the FM-S interface
$P_{n0}/P_F$ for two polarizations of a ferromagnet $P_F=0.95$ (top
panel) and $P_F=0.5$ (bottom panel) 
for two values of the dimensionless parameter $J_s/J_m=5$ and $0.2$. 
The broken curve is for $P_{n0}$ for the case where the density
dependence of the diffusion coefficient has been neglected
($\bar{D}=D_0$). Although 
the absolute values of accumulated polarization are very different,
the relative polarizations, $P_{n0}/P_F$ are almost independent of $P_F$. 
}
\label{fig:fig4}
\end{figure}

General analytical solution for polarization $P_{n0}$ is readily found by
noticing that the right hand side of Eq.~(\ref{eq:bclin}) is very close to a
linear function of $P_{n0}$ at all values of parameters. Therefore, a
general solution for the polarization in degenerate semiconductor can be
written very accurately as 
\begin{eqnarray}
P_{n0} &=&-P_{F}\frac{3J}{3J_{S}L/L_{s}+5J_{m}}  \nonumber \\
&=&-P_{F}\frac{6J}{3\left( \sqrt{J^{2}+4J_{S}^{2}}-J\right) +10J_{m}}.
\label{eq:Pn0J}
\end{eqnarray}

As expected, $P_{n0}$ vanishes with current, $P_{n0}\propto J$ when $|J|<J_{S%
}$, and increases in absolute value when the current approaches
maximum. It follows from Eq.~(\ref{eq:Pn0J}) that the polarization can reach
an absolute maximum only when $J_{m}\gg J_{S}$, where the maximal {\em %
injected} polarization is $P_{n0}(-J_{m})=\frac{3}{8}P_{F},$ and the maximal 
{\em extracted} polarization is $P_{n0}(J_{m})=-0.6P_{F}.$

The solution for $P_{n0}$ (\ref{eq:Pn0J}) together with the expression for
current (\ref{eq:Jsbc1})\ allows one to obtain the I-V curve. The closed
expression for the I-V curve can be obtained in the case of small bias
voltage $|qV|\lesssim \mu _{S},$ where $|P_{n0}|\ll 1:$ 
\begin{equation}
J=\frac{J_{m0}(5J_{m}+3J_{S}L/L_{s})}{5J_{m0}+3J_{S}L/L_{s}}\left[ 1-\left(
1-\frac{qV}{\mu _{S}}\right) ^{5/2}\right] .  \label{C-V}
\end{equation}
where according to (\ref{Jm}) $J_{m0}$ and $J_{m}$ depend
on $V$. The equation (\ref{C-V}) is a transcendental one, but at $|qV|\ll
\mu _{S}$ when $|J|\ll J_{S}$ we have $L/L_{s}\approx 1,$ and it becomes and
an expression for the current that starts as an Ohmic law 
\begin{equation}
J=\frac{5qJ_{m0}(5J_{m}+3J_{S}L/L_{s})}{2\mu _{S}(5J_{m0}+3J_{S}L/L_{s})}V,
\label{OM}
\end{equation}
and then deviates from it at larger bias.

The behavior of injection coefficient $\Gamma $ is very different compared
to the polarization. Using the relation (\ref{eq:Jsx}) for $u=1$\
(neglecting the polarization dependence of the diffusion coefficient)\ and
Eqs.~(\ref{eq:Pn0J}),(\ref{eq:LsJs}), (\ref{eq:LoLs})\ we find a relation
between the injection coefficient $\Gamma $ (polarization of current)\ and
the polarization of density $P_{n}$

\begin{eqnarray}
\Gamma &\equiv &\frac{J_{\uparrow }-J_{\downarrow }}{J}=-P_{n0}\frac{J_{s}L}{%
JL_{s}}  \nonumber \\
&=&3P_{F}\frac{\sqrt{J^{2}+4J_{S`}^{2}}-J}{3\left( \sqrt{J^{2}+4J_{S`}^{2}}%
-J\right) +10J_{m}}.  \label{eq:PJ0}
\end{eqnarray}
The injection coefficient does not vanish with current, but tends to a
finite value 
\begin{equation}
\Gamma =\frac{3P_{F}J_{S}}{3J_{S}+5J_{m}}\text{ when }J=0.
\end{equation}
In order to maximize the polarization $P_{n0}$, according to Eq.~(\ref
{eq:Pn0J}), one needs to use the modified Schottky contact with $J_{m}\gg
J_{S}$ (transparent for tunneling electrons), where the injection
coefficient would be very small, $\Gamma (J\rightarrow 0)\approx
0.6P_{F}J_{S}/J_{m}\ll P_{F}.$ In this case, at large forward current $%
\Gamma (J_{m})=\frac{3}{5}P_{F}(J_{S}/J_{m})^{2}\ll 1,$ so the spin
injection coefficient practically vanishes in spin {\em extraction} regime
(in other words, the polarization of {\em current} vanishes). Very
differently, under reverse bias voltage $\Gamma (-J_{m})=P_{n0}=\frac{3}{8}%
P_{F},$ so that the polarization of the {\em injected} current is large.
Note that here we still assume that the densities of carriers in the FM\
metal and the degenerate semiconductor are vastly different, so that there
is a clear ``conductivity mismatch'' and yet the spin injection proceeds
very efficiently. On the other hand, if we were to make the contact opaque,
where $J_{m}\ll J_{S},$ the polarization of electrons, according to Eq.~(\ref
{eq:Pn0J}), would become minute, $P_{n0}\ll 1,$ since the current through
the structure becomes very small compared to the characteristic current that
polarizes electrons, $J_{S}$. But at the same time, the injection
coefficient becomes large, $\Gamma =P_{F}.$ This is the same behavior as
observed in FM-I-FM tunnel junctions \cite{Brat}: relatively thick tunnel
barriers facilitate strong polarization of {\em current} but the accumulated
spin polarization remains very small since the current density is
insufficient.

The described behavior of the polarization and the injection coefficient is
very important for proper understanding of the behavior of spintronic
structures. In particular, we have demonstrated once again an ill-conceived
nature of the \textquotedblleft conductivity mismatch\textquotedblright\
problem \cite{Sch00}. The condition of the maximum spin accumulation in
semiconductor, $J_{m}\gg J_{S}$, in accordance with Eqs.~(\ref{Jm}), (\ref
{eq:Im}) and (\ref{eq:LsJs}) can be written down as 
\begin{equation}
\alpha _{0}(1-P_{F}^{2})(d_{\uparrow 0}+d_{\downarrow 0})\frac{\tau _{s}v_{T}%
}{L_{s}}\exp \left( -\frac{\eta l}{l_{0}}\right) \gg 1.  \label{eq:barcon}
\end{equation}
This condition can be rewritten with the use of Eq.~(\ref{C-V}) at $qV\ll
\mu _{S}$ as 
\begin{equation}
r_{c}\ll L_{s}/\sigma _{S},  \label{eq:rc}
\end{equation}
where $r_{c}=(dJ/dV)^{-1}$is the tunneling contact resistance. We emphasize
that Eq.~(\ref{eq:rc}) is opposite to the condition of maximum of current
spin polarization found in Ref. \cite{Rash} for small currents. At $r_{c}\gg
L_{s}/\sigma _{S}$, i.e. when $J_{S}\gg J_{m}$, as we noted above, a degree
of spin accumulation in the semiconductor is very small, $P_{n}\ll 1$, but
exactly this $P_{n}$ is the characteristic that determines chief spin
effects \cite{Wolf,Datta,Hot,OB}. Note that the condition (\ref{eq:rc}) does
not depend on the electron concentration, therefore it coincides with that
for nondegenerate semiconductors (see \cite{OB}).

\section{Discussion}

We obtained an analytical solution for spin injection/extraction for
degenerate semiconductor in addition to numerical results for nonlinear spin
diffusion. The nonlinear dependence of the bi-spin diffusion coefficient in
semiconductor on accumulated polarization appears to be small. We emphasize
that the value of $P_{F}$ (\ref{PF}) determining maximum spin polarization
of the FM-S junction depends on bias voltage $V$,\ because the spin factor $%
d_{\sigma }$ given by Eq. (\ref{ds}) is determined by $v_{\sigma
0}=v_{\sigma }(F+qV)$. Since usually $v_{\sigma 0}>v_{t0},$ the spin factor $%
d_{\sigma }$ $\propto v_{\sigma 0}^{-1}$. In a metal, as a rule, $v_{\sigma
0}^{-1}\propto g_{\sigma 0}=g_{\sigma }(F+qV),$ therefore $d_{\sigma
}\propto g_{\sigma }(F+qV),$ where $g_{\sigma 0}=g_{\sigma }(F+qV)$ is the
density of states of the d-electrons with spin $\sigma $ and energy $E=F+qV$
in the ferromagnet. Thus, assuming $m_{\sigma }=m$ we find from Eq. (\ref{PF}%
) that $P_{F}\approx (g_{\uparrow 0}-g_{\downarrow 0})/(g_{\uparrow
0}+g_{\downarrow 0})$. The polarization of d-electrons in elemental
ferromagnets Ni{\it ,} Co,{\it \ }and\ Fe is reduced by the current of
unpolarized s-electrons $rJ_{s}$, where $r<1$ is a factor (roughly the ratio
of the number of s-bands to the number of d-bands crossing the Fermi level).
Together with the contribution of s-electrons the polarization parameter $%
P_{F}$ is approximately 
\begin{equation}
P_{F}=\frac{J_{\uparrow 0}-J_{\downarrow 0}}{J_{\uparrow 0}+J_{\downarrow
0}+J_{s0}}\simeq \frac{g_{\uparrow 0}-g_{\downarrow 0}}{g_{\uparrow
0}+g_{\downarrow 0}+2rg_{s0}},  \label{Pol}
\end{equation}
We note that such a relation for $P_{F}$ can be obtained from a standard
``golden-rule'' type approximation for tunneling current that is supposed
to be proportional to the density of states\ $g_{\sigma }(E)$ (cf. Refs. 
\cite{sze,other,Esaki}). The density of states $g_{\downarrow }$ for {\em %
minority} d-electrons in Fe, Co, and Ni has\ a large peak at $E=E_{F}+\Delta
_{\downarrow }$ ($\Delta _{\downarrow }\simeq 0.1$ eV), much larger than $%
g_{\uparrow }$ for the majority $d-$electrons and $g_{s}$ for $s-$electrons 
\cite{Mor,Maz}, Fig.~2. Therefore, the spin polarization and spin injection
coefficient can potentially achieve a large value of $\left| P_{F}\right| $
in the forward-biased FM-S at a bias voltage $qV=\Delta _{\downarrow }$
(Fig.~2). In reverse biased junctions the situation is different in that
most effective tunneling is by electrons in FM with energies close to the
Fermi level, $E\simeq F$ where the polarization of carriers is positive, $%
P_{F}=40-50\%$ \cite{Maz} and a good fraction of it may be injected into a
semiconductor. In this case the excess of majority spins may be
created in semiconductor for both reverse (injection) and forward
(extraction) bias voltages. This implies a complex dependence of
accumulated spin polarization  on a bias voltage. 

\section{Conclusion}

Let us compare spin injection in the modified Schottky FM-S junctions with
degenerate semiconductor at low temperatures with nondegenerate
semiconductors at large (room) temperature. In both cases the process of
spin injection/extraction strongly depends on current density and is
generally nonlinear. The condition for most efficient spin accumulation is
similar in both cases, $J_{m}\gg J_{S},$ that sets constraints on materials
parameters, see Eq.~(\ref{eq:barcon}). We have studied this case for both
reverse \cite{OB} and forward bias voltages \cite{BOB}.

We have shown that the spin injection in reverse-biased FM-S junctions
differs from that in the forward-biased junctions. In the reverse-biased
junctions spin polarization of injected electrons, $P_{n0}$, and spin
injection coefficient, $\Gamma $, increase with current up to a maximum $%
P_{n0}(-J_{m})=\Gamma (-J_{m})=\frac{3}{8}P_{F},$ where $P_{F}$ is the
polarization of ferromagnet, Fig.~4. In forward-biased FM-S junctions the
polarization approaches $P_{n0}(J_{m})=-0.6P_{F}$ at large currents in a
shrinking region with the width $L\propto 1/J.$ In this case the spin
injection coefficient is small, $\Gamma =0.6P_{F}J_{S}/J_{m}\ll 1$ already
at $J=0,$ and decreases at large current, $\Gamma
(J_{m})=0.6P_{F}(J_{S}/J_{m})^{2}\ll 1.$ Analogous results are obtained in
Refs.~\cite{OB,BOB} for the FM-S junctions with nondegenerate
semiconductors, with the only difference that $\Gamma =P_{F}$ at forward
current when the minority electrons are extracted into the energy region
with a peak in the density of states. The I-V characteristics for Schottky
contacts with degenerate and non-degenerate semiconductors are also quite
different.

It is worth mentioning a different dependence of effective polarization of
ferromagnet $P_{F}$ (\ref{PF}) on bias voltage in both cases. In a
nondegenerate semiconductor $P_{F}$ corresponds to the electron energy $%
E=E_{c0}+qV>F,$ for a degenerate semiconductor it is $E=F+qV$. The value of 
$\left| P_{F}\right| $ for the nondegenerate semiconductor can reach its
maximum at a reverse bias voltage $qV\simeq E_{c0}-F-$ $\Delta _{\downarrow
} $ ($\Delta _{\downarrow }\simeq 0.1$ eV) while for degenerate
semiconductor $P_{F}$ can have the same large value at a forward voltage $%
qV\simeq $ $\Delta _{\downarrow }$, Fig.~2. In a nondegenerate
semiconductor, $P_{F}$ has the same sign practically independently of the
bias while in the degenerate semiconductor $P_{F}$ may change sign with bias
voltage and, at least potentially, can become close to unity at the forward
bias $qV\simeq \Delta _{\downarrow }$, but not at a reverse bias. Under
reverse bias voltage, the electrons are injected into the degenerate
semiconductor from states in the ferromagnet with energies $E\simeq F,$
where their polarization is $\simeq 40-50\%.$ The predicted strong
dependence of accumulated polarization on bias voltage can be exploited in
order to reveal possible effect of peaks in the density of states. Indeed,
if the filling of the conduction band of the degenerate semiconductor is
relatively small, $\mu _{S}\ll 0.1$ eV$\sim \Delta _{\downarrow }$, as is
usually the case, then by changing the forward bias one could ``scan''\ the
density of states in FM with a ``resolution''\ $\mu _{S}\ll \Delta
_{\downarrow },$ Fig.~2. One may see an increase in current and a maximum in
an extracted polarization at $qV=\Delta _{\downarrow }.$ Interestingly, one
may expect a sign change of the polarization in FM-S modified Schottky
contact with a degenerate semiconductor S.

\appendix 
\section{Electroneutrality}

A deviation from the quasineutrality is determined by the continuity
equations (\ref{CC})\ with the Poisson equation (in CGS\ units)\ 
\begin{equation}
dE/dx=-q\Delta n/4\pi \epsilon ,  \label{Pois}
\end{equation}
where $\epsilon $ is the dielectric constant of the material and $\Delta
n=n_{\uparrow }+n_{\downarrow }-n$ is the deviation of electron density from
equilibrium one. We show here that $\Delta n\ll n_{\uparrow },n_{\downarrow
} $ and, therefore, $\Delta n$ can be neglected. To this end, we substitute
the expression for the electric field, Eq.(\ref{Ex}) into (\ref{Pois}) and
obtain the following estimate 
\begin{eqnarray}
\Delta n &=&\frac{\epsilon }{4\pi q}\frac{dE}{dx}=\epsilon \frac{%
(D_{\uparrow }-D_{\downarrow })}{4\pi \sigma }\frac{d^{2}\delta n_{\uparrow }%
}{dx^{2}},  \nonumber \\
\left| \Delta n\right| &\sim &\frac{\epsilon D_{0}}{4\pi \sigma }\frac{%
\delta n_{\uparrow }}{L_{s}^{2}}\sim n\left( \frac{L_{TF}}{L_{s}}\right)
^{2},  \label{eq:Deltan}
\end{eqnarray}
where $L_{TF}=\left( \epsilon D_{0}/4\pi \sigma \right) ^{1/2}=\left(
\epsilon /4\pi g_{F}\right) ^{1/2}$ is the screening length in the
degenerate semiconductor (Thomas-Fermi length). We have used the Einstein
relation $\sigma =q^{2}Dg_{F},$ where $g_{F}$ is the density of states at
the Fermi level. Finally, the required estimate for deviations from
electroneutrality becomes 
\begin{equation}
\frac{\left| \Delta n\right| }{n}\sim \left( \frac{L_{TF}}{L_{s}}\right)
^{2}\ll 1.
\end{equation}
For example, in Si at doping $n\sim 10^{17}$ cm$^{-3}$ the screening length
is $L_{TF}\sim 30$\AA\ and with $L_{s}\sim 1\mu $m one obtains $\Delta
n/n\sim 10^{-5}$, a very small deviation from electroneutrality indeed{\em \ 
}that can be safely neglected (cf. this with attempts to account for
deviations from electroneutrality in Ref. \cite{zuticRev04}).

\section{ Quasi-Fermi level splitting in ferromagnet and
semiconductor}

Let us prove that we indeed can neglect the splitting of the quasi-Fermi
level in FM\ metal, $\Delta F_{\sigma 0}^{f}$, compared to the splitting of
the quasi-Fermi level in semiconductor $\Delta F_{\sigma 0}^{S}$ for the
FM-S junction under consideration. Since we can neglect the electric field
in FM metal, the distribution of spin polarized electrons is determined by
their diffusion: $\delta n_{\uparrow }^{f}(x)=\delta n_{\uparrow 0}^{f}\exp
(x/L_{f})$ in FM, i.e. in the region corresponding to $x<0$, Fig.~1. Thus,
according to (\ref{J1}) the currents of spin polarized electrons in FM near
FM-S interface are 
\begin{equation}
J_{\uparrow (\downarrow )0}=\frac{\sigma _{\uparrow (\downarrow )0}^{f}}{%
\sigma }J+(-)\frac{qL_{f}}{\tau _{s}}\delta n_{\uparrow 0}^{f}.  \label{cfm}
\end{equation}
We find from (\ref{cfm}) and (\ref{PJ}) that $J_{\uparrow 0}\equiv (1+\Gamma
)J/2=(1+P_{FM})J/2+qL_{f}\delta n_{\uparrow 0}^{f}/\tau _{s}^{f},$ which
gives 
\begin{equation}
\delta n_{\uparrow 0}^{f}=J(\Gamma -P_{FM})\tau _{s}^{f}/2qL_{f},
\label{nfm}
\end{equation}
where we have introduced 
\begin{equation}
P_{FM}=(\sigma _{\uparrow }^{f}-\sigma _{\downarrow }^{f})/\sigma ^{f},
\label{FM}
\end{equation}
the spin polarization of a current in the FM bulk. Thus, if we were to make
the same assumption as Aronov and Pikus in Ref.~\cite{Aron} that $\Gamma
=P_{FM}$, we would have obtained $\delta n_{\uparrow 0}^{f}=0$ and,
consequently, $\Delta F_{\sigma 0}^{f}=0$. In other words, there would be no
splitting at all of the Fermi levels in a ferromagnet in Aronov-Pikus
approximation. In reality, there is a splitting of the quasi-Fermi levels in
FM, but it is usually small compared to the splitting in the semiconductor
(see estimates below). This allows us to considerably simplify the
description of the spin injection/extraction.

According to Eqs.~(\ref{nfm}) and (\ref{eq:PJ0}), 
\begin{equation}
\left| \frac{\delta n_{\uparrow 0}^{f}}{\delta n_{\uparrow 0}}\right|
=\gamma \left| \frac{\Gamma -P_{FM}}{\Gamma }\right| ,  \label{rel}
\end{equation}
where $\gamma =\tau _{s}^{f}L_{s}/(\tau _{s}L_{f})\sim 1$. We have shown
that the polarization of current $|\Gamma |<|P_{FM}|,$ and it may be $\ll
|P_{FM}|,$ therefore, $|\delta n_{\uparrow 0}^{f}/\delta n_{\uparrow
0}|\lesssim 1.$

The ratio of $\Delta F_{\uparrow 0}^{S}$ and $\Delta F_{\uparrow 0}^{f}$ is
approximately equal to 
\begin{equation}
\left| \frac{\Delta F_{\uparrow 0}^{f}}{\Delta F_{\uparrow 0}^{S}}\right|
\simeq \frac{n\mu ^{f}}{n^{f}\mu _{S}}\left| \frac{\delta n_{\uparrow 0}^{f}%
}{\delta n_{\uparrow 0}}\right| =\gamma \beta \left| \frac{\Gamma -P_{FM}}{%
\Gamma }\right| \ll 1,  \label{FF}
\end{equation}
where $\beta =n\mu ^{f}/(n^{f}\mu _{S})$, $\mu ^{f}$ and $\mu _{S}$ are the
Fermi energies for electrons in FM and S, respectively. Since the electron
density in FM metals, $n^{f}\simeq 10^{22}$cm$^{-3}$, is several orders of
magnitude larger than in S (typically, $n\lesssim 10^{18}$cm$^{-3}$), the
value $\beta \sim (n/n^{f})^{1/3}\ll 1$. Thus, one can see from (\ref{FF})
that indeed $\left| \Delta F_{\uparrow 0}^{f}\right| \ll \left| \Delta
F_{\uparrow 0}^{S}\right| $. We showed above (see also Ref. \cite{BOB}) that 
$\Gamma $ is small at very large forward currents, therefore, $\Delta
F_{\uparrow 0}^{f}$ can be on the order of $\Delta F_{\uparrow 0}^{S}$.
However, such current corresponds to the bias voltages of Schottky junctions 
$qV\gg \Delta F_{\sigma 0}^{f}$. Due to the condition $\Delta F_{\sigma
0}^{f}\ll \Delta F_{\sigma 0}^{S},q\left| V\right| ,$ we can indeed neglect $%
\Delta F_{\sigma 0}^{f}$ in Eq.~(\ref{eq:Jsbc}), so the approximation used
to derive the Eqs. (\ref{eq:Jsbc}) and~(\ref{eq:bcmain}) is justified.

We emphasize that the conclusion $\Delta F_{\uparrow 0}^{f}\ll \Delta
F_{\uparrow 0}^{S}$ is valid for FM-S Schottky junctions. Possible exception
can only be the FM-S junctions with an accumulation layer, Fig.~1d. In such
FM-S junctions the electron density in S\ near the FM-S interface can be
very large. In such rare case both $\beta $ and $\gamma $ can be on the
order of unity, perhaps allowing for $\Delta F_{\uparrow 0}^{f}\simeq \Delta
F_{\uparrow 0}^{S}$ and even $\zeta _{\sigma 0}^{S}\simeq \zeta _{\sigma
0}^{f}$. However, this case requires a separate study where one has also to
take into account spin selective properties of such a FM-S junctions and a
steep spatial variation of electron density in S near the FM-S interface,
Fig.~1d.

\end{document}